\documentclass{article}
\usepackage{amsmath}
\usepackage{amssymb}
\usepackage{amsfonts}

\input{tcilatex}

\begin{document}

\title{Stochastic solutions of nonlinear PDE's and an extension of
superprocesses}
\author{R. Vilela Mendes \thanks{%
vilela@cii.fc.ul.pt, rvilela.mendes@gmail.com} \\
CMAF - Instituto de Invesrtiga\c{c}\~{a}o Interdisciplinar UL \\
(Av. Gama Pinto 2, 1649-003, Lisbon)\\
Instituto de Plasmas e Fus\~{a}o Nuclear, IST \\
(Av. Rovisco Pais, 1049-001 Lisbon)}
\date{ }
\maketitle

\begin{abstract}
Stochastic solutions provide new rigorous results for nonlinear PDE's and,
through its local non-grid nature, are a natural tool for parallel
computation. There are two different approaches for the construction of
stochastic solutions: MacKean's and superprocesses. However, when restricted
to measures, superprocesses can only be used to generate solutions for a
limited class of nonlinear PDE's. A new class of superprocesses, namely
superprocesses on signed measures and on distributions, is proposed to
extend the stochastic solution approach to a wider class of PDE's.
\end{abstract}

\section{Introduction}

A stochastic solution of a linear or nonlinear partial differential equation
is a stochastic process which, when started from a particular point $x$ in
the domain generates after time $t$ a boundary measure which, when
integrated over the initial condition at $t=0$, provides the solution at the
point $x$ and time $t$. For example for the heat equation%
\begin{equation}
\partial _{t}u(t,x)=\frac{1}{2}\frac{\partial ^{2}}{\partial x^{2}}%
u(t,x)\qquad \text{\textnormal{with}}\qquad u(0,x)=f(x)  \label{1.1}
\end{equation}%
the stochastic process is Brownian motion and the solution is 
\begin{equation}
u(t,x)=\mathbb{E}_{x}f(X_{t})  \label{1.2}
\end{equation}%
$\mathbb{E}_{x}$ meaning the expectation value, starting from $x$, of the
process%
\begin{equation}
dX_{t}=dB_{t}  \label{1.3}
\end{equation}%
The domain here is $\mathbb{R}\times \left[ 0,t\right) $ and the expectation
value in (\ref{1.2}) is indeed the inner product $\left\langle \mu
_{t},f\right\rangle $ of the initial condition $f$ with the measure $\mu
_{t} $ generated by the Brownian motion at the $t-$boundary. The usual
integral solution,%
\begin{equation}
u\left( t,x\right) =\frac{1}{2\sqrt{\pi }}\int \frac{1}{\sqrt{t}}\exp \left(
-\frac{\left( x-y\right) ^{2}}{4t}\right) f\left( y\right) dy  \label{1.4}
\end{equation}%
with the heat kernel, has exactly the same interpretation. Of course, an
important condition for the stochastic process (Brownian motion in this
case) to be considered \textit{the} solution of the equation is the fact
that the same process works for any initial condition. This should be
contrasted with stochastic processes constructed from particular solutions.

That the solutions of linear elliptic and parabolic equations, both with
Cauchy and Dirichlet boundary conditions, have a probabilistic
interpretation is a classical result and a standard tool in potential theory 
\cite{Getoor} \cite{Bass1} \cite{Bass2}. In contrast with the linear
problems, explicit solutions in terms of elementary functions or integrals
for nonlinear partial differential equations are only known in very
particular cases. Therefore the construction of solutions through stochastic
processes, for nonlinear equations, has become an active field in recent
years. The first stochastic solution for a nonlinear pde was constructed by
McKean \cite{McKean} for the KPP equation. Later on, the exit measures
provided by diffusion plus branching processes \cite{Dynkin1} \cite{Dynkin2}
as well as the stochastic representations recently constructed for the
Navier-Stokes \cite{Jan} \cite{Waymire} \cite{Bhatta1} \cite{Ossiander} \cite%
{Orum}, the Vlasov-Poisson \cite{Vilela1} \cite{Vilela2} \cite{Vilela4}, the
Euler \cite{Vilela3} and a fractional version of the KPP equation \cite%
{Cipriano} define solution-independent processes for which the mean values
of some functionals are solutions to these equations. Therefore, they are 
\textit{exact stochastic solutions}.

In the stochastic solutions one deals with a process that starts from the
point where the solution is to be found, a functional being then computed on
the boundary or in some cases along the whole sample path. In addition to
providing new exact results, the stochastic solutions are also a promising
tool for numerical implementation, in particular for parallel computing
using the recently develop probabilistic domain decomposition method \cite%
{Acebron1} \cite{Acebron2} \cite{Acebron3}. This method decomposes the space
in subdomains and then uses in each one a deterministic algorithm with
Dirichlet boundary conditions, the values on the boundaries being determined
by a stochastic algorithm. This minimizes the time-consuming communication
problem between domains and allows for extraordinary improvements in
computer time.

There are basically two methods to construct stochastic solutions. The first
method, which will be called the McKean method, is essentially a
probabilistic interpretation of the Picard series. The differential equation
is written as an integral equation which is rearranged in a such a way that
the coefficients of the successive terms in the Picard iteration obey a
normalization condition. The Picard iteration is then interpreted as an
evolution and branching process, the stochastic solution being equivalent to
importance sampling of the normalized Picard series. The second method
constructs the boundary measures of a measure-valued stochastic process (a
superprocess) and obtains the solution of the differential equation by a
scaling procedure.

For a detailed comparison of the two methods refer to \cite{Vilela5}. Here
after a short review of the construction of superprocesses one illustrates
the basic reason why this construction, when restricted to measure-valued
superprocesses, can only be applied to a limited class of nonlinear partial
differential equations. A wider class of superprocesses, namely
superprocesses on signed measures and on distributions might provide useful
stochastic solutions for some other nonlinear PDE's.

\section{The construction of superprocesses}

In the past, superprocesses have been constructed in the space $M_{+}\left(
E\right) $ of finite measures on a measurable space $\left( E,\mathcal{B}%
\right) $. Because this setting somehow restricts the range of nonlinear
equations for which solutions may be obtained by superprocesses, it is
convenient to use a wider framework.

Let $\mathcal{S}$ be the Schwartz space of functions of rapid decrease and $%
\mathcal{U}\subset \mathcal{S}$ those functions in $\mathcal{S}$ that may be
extended into the complex plane as entire functions of rapid decrease on
strips. $\mathcal{U}^{\prime }$, the dual of $\mathcal{U}$, is Silva's space
of tempered ultradistributions \cite{Silva1} \cite{Silva2}, which can also
be characterized as the space of all Fourier transforms of distributions of
exponential type \cite{Hoskins}. Furthermore, for reasons to be clear later
on, it is convenient to restrict oneself to the space $\mathcal{U}%
_{0}^{\prime }$ of tempered ultradistributions of compact support \cite%
{Silva2}.

Denote now by $\left( X_{t},P_{0\,,\nu }\right) $ a branching stochastic
process with values in $\mathcal{U}_{0}^{\prime }$ and transition
probability $P_{0,\nu }$ starting from time $0$ and $\nu \in \mathcal{U}%
_{0}^{\prime }$. The process is said to satisfy the \textit{branching
property} if given $\nu =\nu _{1}+\nu _{2}$%
\begin{equation}
P_{0,\nu }=P_{0,\nu _{1}}\ast P_{0,\nu _{2}}  \label{2.5}
\end{equation}%
that is, after the branching $\left( X_{t}^{1},P_{0,\nu _{1}}\right) $ and $%
\left( X_{t}^{2},P_{0\,,\nu _{2}}\right) $ are independent and $%
X_{t}^{1}+X_{t}^{2}$ has the same law as $\left( X_{t},P_{0,\nu }\right) $.
In terms of the \textit{transition operator }$V_{t}$ operating on functions
on $\mathcal{U}$ this is%
\begin{equation}
\left\langle V_{t}f,\nu _{1}+\nu _{2}\right\rangle =\left\langle V_{t}f,\nu
_{1}\right\rangle +\left\langle V_{t}f,\nu _{2}\right\rangle  \label{2.6}
\end{equation}%
where $e^{-\left\langle V_{t}f,\nu \right\rangle }\circeq P_{0,\nu
}e^{-\left\langle f,X_{t}\right\rangle }$ or%
\begin{equation}
\left\langle V_{t}f,\nu \right\rangle =-\log P_{0,\nu }e^{-\left\langle
f,X_{t}\right\rangle }  \label{2.7}
\end{equation}%
$f\in \mathcal{U},\nu \in \mathcal{U}_{0}^{\prime }$.

Underlying the usual construction of superprocesses, in the form that is
useful for the representation of solutions of PDE's, there is a stochastic
process with paths that start from a particular point in $E$, then propagate
and branch but the paths preserve the same nature after the branching. In
terms of measures it means that one starts from an initial $\delta _{x}$
which at branching originate other $\delta ^{\prime }s$ with at most some
scaling factors. It is to preserve this pointwise interpretation that, in
this larger setting, we are restricting to ultradistributions in $\mathcal{U}%
_{0}^{\prime }$. Any ultradistribution in $\mathcal{U}_{0}^{\prime }$ may be
represented as a multipole expansion at any point of its support, that is as
a series of $\delta ^{\prime }s$ and their derivatives \cite{Silva2}.
Therefore any arbitrary transition in the process $X_{t}$ in $\mathcal{U}%
_{0}^{\prime }$ may be associated to a branching of paths in $E$ and along
these new paths new distributions with point support propagate. As a result
the construction now proceeds as in the measure-valued case.

In $M=\left[ 0,\infty \right) \times E$ consider a set $Q\subset M$ and the
associated exit process $\xi =\left( \xi _{t},\Pi _{0,x}\right) $ with
parameter $k$ defining the lifetime. The process stars from $x\in E$
carrying along an ultradistribution in $\mathcal{U}_{0}^{\prime }$ with
support on the path. At each branching point of the $\xi _{t}-$process there
is a transition ruled by the $P$ probability in $\mathcal{U}_{0}^{\prime }$
leading to one or more elements in $\mathcal{U}_{0}^{\prime }$. These $%
\mathcal{U}_{0}^{\prime }$ elements are then carried along by the new paths
of the $\xi _{t}-$process. The whole process stops at the boundary $\partial
Q$, finally defining a exit process $\left( X_{Q},P_{0,\nu }\right) $ on $%
\mathcal{U}_{0}^{\prime }$. If the initial $\nu $ is $\delta _{x}$ one writes%
\begin{equation}
u\left( x\right) =\left\langle V_{Q}f,\nu \right\rangle =-\log
P_{0,x}e^{-\left\langle f,X_{Q}\right\rangle }  \label{2.10}
\end{equation}%
$\left\langle f,X_{Q}\right\rangle $ being computed on the (space-time)
boundary with the exit ultradistribution generated by the process.

The connection with nonlinear PDE's is established by defining the whole
process to be a $\left( \xi ,\psi \right) -$\textit{superprocess} if $%
u\left( x\right) $ satisfies the equation%
\begin{equation}
u+G_{Q}\psi \left( u\right) =K_{Q}f  \label{2.11}
\end{equation}%
where $G_{Q}$ is the Green operator,%
\begin{equation}
G_{Q}f\left( r,x\right) =\Pi _{0,x}\int_{0}^{\tau }f\left( s,\xi _{s}\right)
ds  \label{2.12}
\end{equation}%
and $K_{Q}$ the Poisson operator%
\begin{equation}
K_{Q}f\left( x\right) =\Pi _{0,x}1_{\tau <\infty }f\left( \xi _{\tau }\right)
\label{2.13}
\end{equation}%
$\psi \left( u\right) $ means $\psi \left( 0,x;u\left( 0,x\right) \right) $
and $\tau $ is the first exit time from $Q$.

The superprocess is constructed as follows: Let $\varphi \left( s,x;z\right) 
$ be the branching function at time $s$ and point $x$. Then for $e^{-w\left(
0,x\right) }\circeq P_{0,x}e^{-\left\langle f,X_{Q}\right\rangle }$ one has%
\begin{equation}
P_{0,x}e^{-\left\langle f,X_{Q}\right\rangle }\circeq e^{-w\left( 0,x\right)
}=\Pi _{0,x}\left[ e^{-k\tau }e^{-f\left( \tau ,\xi _{\tau }\right)
}+\int_{0}^{\tau }dske^{-ks}\varphi \left( s,\xi _{s};e^{-w\left( \tau
-s,\xi _{s}\right) }\right) \right]  \label{2.15}
\end{equation}%
$\tau $ is the first exit time from $Q$ and $f\left( \tau ,\xi _{\tau
}\right) =\left\langle f,X_{Q}\right\rangle $ is computed with the exit
boundary ultradistribution. For measure-valued superprocesses the branching
function would be 
\begin{equation}
\varphi \left( s,y;z\right) =c\sum_{0}^{\infty }p_{n}(s,y)z^{n}  \label{2.14}
\end{equation}%
with $\sum_{n}p_{n}=1$ and $c$ denoting the branching intensity, but now it
may be a much more general function.

For the interpretation of the superprocesses as generating solutions of
PDE's, an essential role is played by a transformation of Eq.(\ref{2.15})
that uses $\int_{0}^{\tau }ke^{-ks}ds=1-e^{-k\tau }$ and the Markov property 
$\Pi _{0,x}1_{s<\tau }\Pi _{s,\xi _{s}}=\Pi _{0,x}1_{s<\tau }$. This is
lemma 1.2 in ch.4 of Ref.\cite{Dynkin1}. Because it only depends on the
Markov properties of the $\left( \xi _{t},\Pi _{0,x}\right) $ process it
also holds in this more general context. A proof is included in the Appendix
with the notations used in this paper.

Using the lemma, Eq.(\ref{2.15}) for $e^{-w\left( 0,x\right) }$ is converted
into%
\begin{equation}
e^{-w\left( 0,x\right) }=\Pi _{0,x}\left[ e^{-f\left( \tau ,\xi _{\tau
}\right) }+k\int_{0}^{\tau }ds\left[ \varphi \left( s,\xi _{s};e^{-w\left(
\tau -s,\xi _{s}\right) }\right) -e^{-w\left( \tau -s,\xi _{s}\right) }%
\right] \right]  \label{2.16}
\end{equation}%
Eq.(\ref{2.11}) is now obtained by a limiting process. Let in (\ref{2.16})
replace $w\left( 0,x\right) $ by $\beta w_{\beta }\left( 0,x\right) $ and $f$
by $\beta f$. $\beta $ is interpreted as the mass of the particles and when
the $\mathcal{U}_{0}^{\prime }$-valued process $X_{Q}\rightarrow \beta X_{Q}$
then $P_{\mu }\rightarrow P_{\frac{\mu }{\beta }}$.%
\begin{equation}
e^{-\beta w\left( 0,x\right) }=\Pi _{0,x}\left[ e^{-\beta f\left( \tau ,\xi
_{\tau }\right) }+k_{\beta }\int_{0}^{\tau }ds\left[ \varphi _{\beta }\left(
s,\xi _{s};e^{-\beta w\left( \tau -s,\xi _{s}\right) }\right) -e^{-\beta
w\left( \tau -s,\xi _{s}\right) }\right] \right]  \label{2.17}
\end{equation}%
Two scaling limits will be used in this paper. The first one, which is the
one used in the past for superprocesses on measures, defines%
\begin{equation}
u_{\beta }^{(1)}=\left( 1-e^{-\beta w_{\beta }}\right) /\beta \hspace{0.3cm};%
\hspace{0.3cm}f_{\beta }^{(1)}=\left( 1-e^{-\beta f}\right) /\beta
\label{2.18}
\end{equation}%
and%
\begin{equation}
\psi _{\beta }^{(1)}\left( 0,x;u_{\beta }^{(1)}\right) =\frac{k_{\beta }}{%
\beta }\left( \varphi \left( 0,x;1-\beta u_{\beta }^{(1)}\right) -1+\beta
u_{\beta }^{(1)}\right)  \label{2.19}
\end{equation}%
one obtains from (\ref{2.17})%
\begin{equation}
u_{\beta }^{(1)}\left( 0,x\right) +\Pi _{0,x}\int_{0}^{\tau }ds\psi _{\beta
}^{(1)}\left( s,\xi _{s};u_{\beta }^{(1)}\right) =\Pi _{0,x}f_{\beta
}^{(1)}\left( \tau ,\xi _{\tau }\right)  \label{2.20}
\end{equation}%
that is%
\begin{equation}
u_{\beta }^{(1)}+G_{Q}\psi _{\beta }^{(1)}\left( u_{\beta }^{(1)}\right)
=K_{Q}f_{\beta }^{(1)}  \label{2.21}
\end{equation}%
When $\beta \rightarrow 0$, $f_{\beta }^{(1)}\rightarrow f$ and if $\psi
_{\beta }$ goes to a well defined limit $\psi $ then $u_{\beta }$ tends to a
limit $u$ solution of (\ref{2.11}) associated to a superprocess. Also one
sees from (\ref{2.18}) that in the $\beta \rightarrow 0$ limit%
\begin{equation}
u_{\beta }^{(1)}\rightarrow w_{\beta }=-\log P_{0,x}e^{-\left\langle
f,X_{Q}\right\rangle }  \label{2.21a}
\end{equation}%
as in Eq.(\ref{2.10}). The superprocess corresponds to a cloud of particles
for which both the mass and the lifetime tend to zero.

\section{Measure-valued superprocesses and nonlinear PDE's}

Here one restricts oneself to measure-valued superprocesses, that is, in
terms of paths, to $\delta ^{\prime }s$ propagating along the paths of the $%
\left( \xi _{t},\Pi _{0,x}\right) $ process and simply branching to new $%
\delta $ measures at each branching point. Let us construct a superprocess
providing a solution to the equation%
\begin{equation}
\frac{\partial u}{\partial t}=\frac{1}{2}\frac{\partial ^{2}u}{\partial x^{2}%
}-u^{\alpha }  \label{2.34}
\end{equation}%
for $1<\alpha \leq 2$. Comparing with (\ref{2.11}) one should have%
\begin{equation*}
\psi \left( 0,x;u\right) =u^{\alpha }
\end{equation*}%
Then from (\ref{2.19}) and (\ref{2.14}), with $z=1-\beta u_{\beta }^{(1)}$
one has%
\begin{eqnarray}
\varphi \left( 0,x;z\right) &=&\sum_{n}p_{n}z^{n}=z+\frac{\beta }{k_{\beta }}%
u_{\beta }^{(1)\alpha }=z+\frac{\beta }{k_{\beta }}\frac{\left( 1-z\right)
^{\alpha }}{\beta ^{\alpha }}  \notag \\
&=&z+\frac{1}{k_{\beta }\beta ^{\alpha -1}}\left( 1-\alpha z+\frac{\alpha
\left( \alpha -1\right) }{2}z^{2}-\frac{\alpha \left( \alpha -1\right)
\left( \alpha -2\right) }{3!}z^{3}+\cdots \right)  \notag \\
&&  \label{2.32}
\end{eqnarray}%
Choosing $k_{\beta }=\frac{\alpha }{\beta ^{\alpha -1}}$ the terms in $z$
cancel and for $1<\alpha \leq 2$ the coefficients of all the remaining $z$
powers are positive and may be interpreted as branching probabilities. It
would not be so for $\alpha >2$. Then%
\begin{equation}
p_{0}=\frac{1}{\alpha };\hspace{0.3cm}p_{1}=0;\hspace{0.3cm}\cdots \hspace{%
0.3cm}p_{n}=\frac{\left( -1\right) ^{n}}{\alpha }\left( 
\begin{array}{l}
\alpha \\ 
n%
\end{array}%
\right) \hspace{0.3cm}n\geq 2  \label{2.33}
\end{equation}%
with $\sum_{n}p_{n}=1$. With this choice of probabilities $p_{n}$ for
branching into new $\delta $ measures and with $k_{\beta }=\frac{\alpha }{%
\beta ^{\alpha -1}}$ and $\beta \rightarrow 0$ one obtains a superprocess
which, through (\ref{2.10}), provides a solution to the Eq.(\ref{2.34}). $%
\alpha =2$ is an upper bound for this representation, because for $\alpha >2$
some of the $p_{n}^{\prime }s$ would be negative and would not be
interpretable as branching probabilities.

For the particular case%
\begin{equation}
\frac{\partial u}{\partial t}=\frac{1}{2}\frac{\partial ^{2}u}{\partial x^{2}%
}-u^{2}  \label{2.28a}
\end{equation}%
\begin{equation}
p_{1}=0;\hspace{0.3cm}p_{0}=p_{2}=\frac{1}{2};\hspace{0.3cm}k_{\beta }=\frac{%
2}{\beta }  \label{2.31}
\end{equation}%
When $\beta \rightarrow 0$, the solutions are given by (\ref{2.10}) and the
superprocesses correspond to the scaling limit of processes where both the
mass and the lifetime of the particles tends to zero and at each bifurcation
point one has probability $p_{0}$ of dying without offspring or creating $n$
new $\delta $ measures with probabilities $p_{n}$.

Superprocesses are usually associated with nonlinear PDE's in the scaling
limit $\beta \rightarrow 0$ of (\ref{2.19})-(\ref{2.20}). However other
limits may also be useful. For example with with $p_{n}=\delta _{n,2}$, $%
\beta =1$ and $k_{\beta }=1$ one obtains%
\begin{eqnarray}
\psi _{\beta }^{(1)}\left( 0,x;u_{\beta }^{(1)}\right) &=&\frac{k_{\beta }}{%
\beta }\left( \varphi \left( 0,x;1-\beta u_{\beta }^{(1)}\right) -1+\beta
u_{\beta }^{(1)}\right)  \notag \\
&=&\frac{k_{\beta }}{\beta }\left( \sum p_{n}\left( 1-\beta u_{\beta
}^{(1)}\right) ^{n}-1+\beta u_{\beta }^{(1)}\right)  \notag \\
&=&\frac{k_{\beta }}{\beta }\left( \beta ^{2}u_{\beta }^{(1)2}-\beta
u_{\beta }^{(1)}\right)  \notag \\
&\rightarrow &u^{2}-u  \label{2.27}
\end{eqnarray}%
Therefore, in this case, one is led to the KPP equation%
\begin{equation}
\frac{\partial u}{\partial t}=\frac{1}{2}\frac{\partial ^{2}u}{\partial x^{2}%
}-u^{2}+u  \label{2.23}
\end{equation}%
However in this case, because $\beta =1$ instead of $\beta \rightarrow 0$,
the solution is given by $\left( 1-e^{-w}\right) $ instead of (\ref{2.10}).
Because of the natural stochastic clock provided by the linear $u$ term, a
stochastic solution for the Cauchy problem of the KPP\ equation may be
constructed by other method \cite{McKean}. However, the interpretation as an
exit measure, allows for the construction of solutions with arbitrary
boundary conditions.

\section{Superprocesses on signed measures and ultradistributions}

Although the scaling limit $\beta \rightarrow 0$ of measure-valued
superprocesses allows the construction of solutions for equations which do
not possess a natural Poisson clock, it has the severe limitation of
requiring a polynomial branching function $\varphi \left( s,x;z\right) $.
This automatically restricts the nonlinear terms in the pde's to be powers
of $u$. In addition, these terms must be such that all coefficients in the $%
z^{n}$ expansion in Eq.(\ref{2.14}) be positive to be interpretable as
branching probabilities. As seen before, it was this requirement that led to
the restriction $1<\alpha \leq 2$ in (\ref{2.34}).

The variable $z$ that appears in $\varphi _{\beta }\left( s,x;z\right) $ is
in fact $z=e^{-\beta w\left( \tau -s,\xi _{s}\right)
}=P_{0,x}e^{-\left\langle \beta f,X\right\rangle }$. When restricting the
superprocess to measures, the delta measure, at each branching point, may at
most branch into other deltas (with positive coefficients) and therefore $%
\varphi \left( s,x;z\right) $ must be a sum of monomials in $z$. When one
generalizes to $\mathcal{U}_{0}^{\prime }$ ultradistributions of point
support\footnote{%
Because distributions of point support are a finite sum of deltas and their
derivatives \cite{Stein}, one could have considered only distributions of
point support rather than compact support ultradistributions $\mathcal{U}%
_{0}^{\prime }$. However in $\mathcal{U}_{0}^{\prime }$ one is not
restricted to finite sums.}, changes of sign and transitions from deltas to
their derivatives are allowed. In the end, the exponential $e^{-\left\langle
\beta f,X\right\rangle }$ will be computed by evaluation of the function on
the ultradistribution that reaches the boundary. To find out the equation
that is represented by the process one needs to compute the $\psi _{\beta
}\left( 0,x;u_{\beta }\right) $ of Eq.(\ref{2.19}) for the corresponding $%
\varphi \left( s,x;z\right) $ in the $\beta \rightarrow 0$ limit. Recalling
that $\varphi \left( s,x;z\right) =\varphi _{\beta }\left( s,\xi
_{s};e^{-\beta w\left( \tau -s,\xi _{s}\right) }\right) $ and $z=e^{-\beta
w_{\beta }}$, one concludes that there are basically two new transitions at
the branching points:

1) A change of sign in the point support ultradistribution%
\begin{equation}
e^{\left\langle \beta f,\delta _{x}\right\rangle }=e^{\beta f\left( x\right)
}\rightarrow e^{\left\langle \beta f,-\delta _{x}\right\rangle }=e^{-\beta
f\left( x\right) }  \label{3.1}
\end{equation}%
which corresponds to%
\begin{equation}
z\rightarrow \frac{1}{z}  \label{3.1a}
\end{equation}%
and

2) A change from $\delta ^{(n)}$ to $\pm \delta ^{(n+1)}$, for example%
\begin{equation}
e^{\left\langle \beta f,\delta _{x}\right\rangle }=e^{\beta f\left( x\right)
}\rightarrow e^{\left\langle \beta f,\pm \delta _{x}^{\prime }\right\rangle
}=e^{\mp \beta f^{\prime }\left( x\right) }  \label{3.2}
\end{equation}%
which corresponds to%
\begin{equation}
z\rightarrow e^{\mp \partial _{x}\log z}  \label{3.2a}
\end{equation}%
Case 1) corresponds to an extension of superprocesses on measures to
superprocesses on signed measures and the second to superprocesses in $%
\mathcal{U}_{0}^{\prime }$.

How these transformations provide stochastic representations of solutions
for other classes of pde's, will be illustrated by two examples:

First, let%
\begin{equation}
\varphi ^{(1)}\left( 0,x;z\right) =p_{1}e^{\partial _{x}\log
z}+p_{2}e^{-\partial _{x}\log z}+p_{3}z^{2}  \label{3.3}
\end{equation}%
This branching function means that at the branching point, with probability $%
p_{1}$ a derivative is added to the propagating ultradistribution, with
probability $p_{2}$ a derivative is added plus a change of sign and with
probability $p_{3}$ the ultradistribution branches into two identical ones.
Using the transformation and scaling limit (\ref{2.18}) one has, for small $%
\beta $%
\begin{equation}
z\rightarrow e^{\mp \partial _{x}\log z}=e^{\mp \partial _{x}\log \left(
1-\beta u_{\beta }^{(1)}\right) }=1\pm \beta \partial _{x}u_{\beta }^{(1)}+%
\frac{\beta ^{2}}{2}\left\{ \left( \partial _{x}u_{\beta }^{(1)}\right)
^{2}\pm \partial _{x}u_{\beta }^{(1)2}\right\} +O\left( \beta ^{3}\right)
\label{3.4a}
\end{equation}%
\begin{equation}
z\rightarrow z^{2}=\left( 1-\beta u_{\beta }^{(1)}\right) ^{2}=1-2\beta
u_{\beta }^{(1)}+\beta ^{2}u_{\beta }^{(1)2}  \label{3.4b}
\end{equation}%
Then, computing $\psi _{\beta }\left( 0,x;u_{\beta }^{(1)}\right) $ with $%
p_{1}=p_{2}=\frac{1}{4}$ and $p_{3}=\frac{1}{2}$ one obtains%
\begin{eqnarray}
\psi _{\beta }^{(1)}\left( 0,x;u_{\beta }^{(1)}\right) &=&\frac{k_{\beta }}{%
\beta }\left( \varphi ^{(1)}\left( 0,x;z\right) -z\right)  \notag \\
&=&\frac{k_{\beta }}{\beta }\left( \varphi ^{(1)}\left( 0,x;1-\beta u_{\beta
}^{(1)}\right) -1+\beta u_{\beta }^{(1)}\right)  \notag \\
&=&\frac{k_{\beta }}{\beta }\left( \frac{1}{8}\beta ^{2}\left( \partial
_{x}u_{\beta }^{(1)}\right) ^{2}+\frac{1}{2}\beta ^{2}u_{\beta
}^{(1)2}+O\left( \beta ^{3}\right) \right)  \label{3.5}
\end{eqnarray}%
meaning that, with $k_{\beta }=\frac{4}{\beta }$, the superprocess provides,
in the $\beta \rightarrow 0$ limit, a solution to the equation%
\begin{equation}
\frac{\partial u}{\partial t}=\frac{1}{2}\frac{\partial ^{2}u}{\partial x^{2}%
}-2u^{2}-\frac{1}{2}\left( \partial _{x}u\right) ^{2}  \label{3.6}
\end{equation}

For the second example a different scaling limit will be used, namely%
\begin{equation}
u_{\beta }^{(2)}=\frac{1}{2\beta }\left( e^{\beta w_{\beta }}-e^{-\beta
w_{\beta }}\right) \hspace{0.3cm};\hspace{0.3cm}f_{\beta }^{(2)}=\frac{1}{%
2\beta }\left( e^{\beta f}-e^{-\beta f}\right)  \label{3.7}
\end{equation}%
Notice that, as before, $u_{\beta }^{(2)}\rightarrow w_{\beta }$ and $%
f_{\beta }^{(2)}\rightarrow f$ when $\beta \rightarrow 0$. In this case with 
$z=e^{\beta w_{\beta }}$ one has%
\begin{eqnarray}
z &=&-2\beta u_{\beta }^{(2)}+2\sqrt{\beta ^{2}u_{\beta }^{(2)2}+1}  \notag
\\
&=&2-2\beta u_{\beta }^{(2)}+\beta ^{2}u_{\beta }^{(2)2}+O\left( \beta
^{4}\right)  \label{3.8a}
\end{eqnarray}%
and%
\begin{eqnarray}
\frac{1}{z} &=&2\beta u_{\beta }^{(2)}+2\sqrt{\beta ^{2}u_{\beta }^{(2)2}+1}
\notag \\
&=&2+2\beta u_{\beta }^{(2)}+\beta ^{2}u_{\beta }^{(2)2}+O\left( \beta
^{4}\right)  \label{3.8b}
\end{eqnarray}%
For the integral equation, instead of (\ref{2.20}), one has%
\begin{equation}
u_{\beta }^{(2)}\left( 0,x\right) +\Pi _{0,x}\int_{0}^{\tau }ds\psi _{\beta
}^{(2)}\left( s,\xi _{s};u_{\beta }^{(2)}\right) =\Pi _{0,x}f_{\beta
}^{(2)}\left( \tau ,\xi _{\tau }\right)  \label{3.9}
\end{equation}%
with%
\begin{equation}
\psi _{\beta }^{(2)}\left( 0,x;u_{\beta }^{(2)}\right) =k_{\beta }\left( 
\frac{1}{2\beta }\left( \varphi \left( 0,x;z\right) -\varphi \left( 0,x;%
\frac{1}{z}\right) \right) -u_{\beta }^{(2)}\right)  \label{3.10}
\end{equation}%
Let now%
\begin{equation}
\varphi ^{(2)}\left( 0,x;z\right) =p_{1}z^{2}+p_{2}\frac{1}{z}  \label{3.11}
\end{equation}%
This branching function means that with probability $p_{1}$ the
ultradistribution branches into two identical ones and with probability $%
p_{2}$ it changes its sign. Therefore, in this case, one is simply extending
the superprocess construction to signed measures. Using (\ref{3.8a}) and (%
\ref{3.8b}) one computes $\psi _{\beta }^{(2)}\left( 0,x;u_{\beta
}^{(2)}\right) $ obtaining%
\begin{equation}
\psi _{\beta }^{(2)}\left( 0,x;u_{\beta }^{(2)}\right) =k_{\beta }\left\{
-p_{1}8u_{\beta }^{(2)}\left( 1+\frac{1}{2}\beta ^{2}u_{\beta
}^{(2)2}\right) +p_{2}u_{\beta }^{(2)}-u_{\beta }^{(2)}+O\left( \beta
^{4}\right) \right\}  \label{3.12}
\end{equation}%
and with $p_{1}=\frac{1}{10};p_{2}=\frac{9}{10}$ and $k_{\beta }=\frac{5}{%
2\beta ^{2}}$ one obtains in the in the $\beta \rightarrow 0$ limit%
\begin{equation}
\psi _{\beta }^{(2)}\left( 0,x;u_{\beta }^{(2)}\right) \rightarrow -u_{\beta
}^{(2)3}  \label{3.13}
\end{equation}%
meaning that this superprocess provides a solution to the equation%
\begin{equation}
\frac{\partial u}{\partial t}=\frac{1}{2}\frac{\partial ^{2}u}{\partial x^{2}%
}+u^{3}  \label{3.14}
\end{equation}

In conclusion: Extending the superprocess construction to signed measures
and ultradistributions, stochastic solutions are obtained for a much larger
class of partial differential equations.

\section{Appendix: Proof of the lemma}

Let%
\begin{equation}
u\left( x,t\right) =\Pi _{0,x}\left\{ e^{-kt}u\left( \xi _{t},0\right)
+\int_{0}^{t}ke^{-ks}\Phi \left( \xi _{s},t-s\right) ds\right\}  \label{A.2}
\end{equation}

Then%
\begin{eqnarray}
\Pi _{0,x}\int_{0}^{t}ku\left( \xi _{s},t-s\right) ds &=&\Pi _{0,x}\left\{
\int_{0}^{t}ke^{-k\left( t-s\right) }u\left( \xi _{s+t-s},0\right) ds\right.
\notag \\
&&\left. +\int_{0}^{t}kds\int_{0}^{t-s}kds^{\prime }e^{-ks^{\prime }}\Phi
\left( \xi _{s+s^{\prime }},t-s-s^{\prime }\right) \right\}  \notag \\
&&  \label{A.3}
\end{eqnarray}%
Summing (\ref{A.2}) and (\ref{A.3})%
\begin{eqnarray}
&&u\left( x,t\right) +\Pi _{0,x}\int_{0}^{t}ku\left( \xi _{s},t-s\right) ds 
\notag \\
&=&\Pi _{0,x}\left\{ \left( e^{-kt}+\int_{0}^{t}ke^{-k\left( t-s\right)
}ds\right) u\left( \xi _{t},0\right) \right.  \notag \\
&&\left. +k\int_{0}^{t}e^{-ks}\Phi \left( \xi _{s},t-s\right)
ds+k\int_{0}^{t}ds\int_{0}^{t-s}kds^{\prime }e^{-ks^{\prime }}\Phi \left(
\xi _{s+s^{\prime }},t-s-s^{\prime }\right) ds^{\prime }\right\}  \notag \\
&&  \label{A.4}
\end{eqnarray}%
Changing variables in the last integral in (\ref{A.4}) from $\left(
s,s^{\prime }\right) $ to $\left( s,\sigma =s+s^{\prime }\right) $ one
obtains for the last term%
\begin{equation*}
k\int_{0}^{t}d\sigma \int_{0}^{\sigma }kdse^{-k\left( \sigma -s\right) }\Phi
\left( \xi _{\sigma },t-\sigma \right) ds
\end{equation*}%
and finally%
\begin{eqnarray}
&&u\left( x,t\right) +\Pi _{0,x}k\int_{0}^{t}u\left( \xi _{s},t-s\right) ds 
\notag \\
&=&\Pi _{0,x}\left\{ u\left( \xi _{t},0\right) +k\int_{0}^{t}\Phi \left( \xi
_{s},t-s\right) ds\right\}  \label{A.5}
\end{eqnarray}


\begin{thebibliography}{99}
\bibitem{Getoor} R. M. Blumenthal and R. K. Getoor; \textit{Markov processes
and potential theory}, Academic Press, New York 1968.

\bibitem{Bass1} R. F. Bass; \textit{Probabilistic techniques in analysis},
Springer, New York 1995.

\bibitem{Bass2} R. F. Bass; \textit{Diffusions and elliptic operators},
Springer, New York 1998.

\bibitem{McKean} H. P. McKean; Comm. on Pure and Appl. Math. 28 (1975)
323-331, 29 (1976) 553-554.

\bibitem{Dynkin1} E. B. Dynkin; \textit{Diffusions, Superdiffusions and
Partial Differential Equations, }AMS Colloquium Pubs., Providence 2002.

\bibitem{Dynkin2} E. B.Dynkin; \textit{Superdiffusions and positive
solutions of nonlinear partial differential equations}, AMS ,
Providence.2004.

\bibitem{Jan} Y. LeJan and A. S. Sznitman ; Prob. Theory and Relat. Fields
109 (1997) 343-366.

\bibitem{Waymire} E. C. Waymire; Prob. Surveys 2 (2005) 1-32.

\bibitem{Bhatta1} R. N. Bhattacharya et al. ; Trans. Amer. Math. Soc. 355
(2003) 5003-5040

\bibitem{Ossiander} M. Ossiander ; Prob. Theory and Relat. Fields 133 (2005)
267-298.

\bibitem{Orum} J. C. Orum; \textit{Stochastic cascades and 2D Fourier
Navier-Stokes equations}, in \textit{Lectures on multiscale and
multiplicative processes,} www.maphysto.dk/publications/MPS-LN/2002/11.pdf

\bibitem{Vilela1} R. Vilela Mendes and F. Cipriano; Commun. Nonlinear
Science and Num. Simul. 13 (2008) 221-226 and 1736.

\bibitem{Vilela2} E. Floriani, R. Lima and R. Vilela Mendes; European
Physical Journal D 46 (2008) 295-302 and 407.

\bibitem{Vilela3} R. Vilela Mendes; Stochastics 81 (2009) 279-297.

\bibitem{Vilela4} R. Vilela Mendes; J. Math. Phys. 51 (2010) 043101.

\bibitem{Cipriano} F. Cipriano, H. Ouerdiane and R. Vilela Mendes; Fract.
Calc. Appl. Anal.\textbf{\ }12 (2009) 47-56.

\bibitem{Acebron1} J. A. Acebr\'{o}n, A. Rodriguez-Rozas and R. Spigler; J.
of Comp. Physics 228 (2009) 5574--5591.

\bibitem{Acebron2} J.A. Acebr\'{o}n, A. Rodr\'{\i}guez-Rozas and R. Spigler;
J. on Scientific Computing 43 (2010) 135-157.

\bibitem{Acebron3} J.A. Acebr\'{o}n and A. Rodr\'{\i}guez-Rozas; J. Comp.
Phys. 230 (2011) 7891-7909.

\bibitem{Vilela5} R. Vilela Mendes; \textit{Stochastic solutions of
nonlinear PDE's: McKean versus superprocesses}, arXiv:1111.5504, in
Proceedings of "Chaos, Complexity and Transport", X. Leoncini (Ed.) World
Scientific 2012.

\bibitem{Silva1} J. Sebasti\~{a}o e Silva; Math. Annalen 136 (1958) 58-96.

\bibitem{Silva2} J. Sebasti\~{a}o e Silva; Math. Annalen 174 (1967) 109-142.

\bibitem{Hoskins} R. F. Hoskins and J. Sousa Pinto; \textit{Distributions,
ultradistributions and other generalized functions}, Ellis Horwood, New York
1994.

\bibitem{Stein} E. M. Stein and R. Shakarchi; \textit{Princeton Lectures in
Analysis IV, Functional Analysis: Introduction to Further Topics in Analysis}%
, Princeton Univ. Press, Princeton 2011.
\end{thebibliography}
\end{document}